\documentclass[pra,onecolumn,floatfix,superscriptaddress,longbibliography,notitlepage, nofootinbib]{revtex4-1}
\pdfoutput=1
\usepackage{tikz}
\usepackage[utf8]{inputenc}
\usepackage{braket}
\usepackage{amsmath}
\DeclareMathOperator{\Tr}{Tr}
\usepackage{dcolumn}   
\usepackage{bm}        
\usepackage{amssymb}
\usepackage{mathtools}
\usepackage{tikz}
\usepackage{qcircuit}
\usepackage{appendix}
\usepackage{hyperref}
\usepackage{subcaption}
\usepackage{amsmath,amsfonts,amssymb}
\usepackage{mathtools}
\usepackage{thmtools,thm-restate}
\usepackage{algorithm}

\usepackage{algpseudocode}

\newtheorem{theorem}{Theorem}
\newtheorem{lemma}{Lemma}

\usetikzlibrary{arrows.meta}

\def\BibTeX{{\rm B\kern-.05em{\sc i\kern-.025em b}\kern-.08em
    T\kern-.1667em\lower.7ex\hbox{E}\kern-.125emX}}

\begin{document}
\title{An Improved Method for Quantum Matrix Multiplication}
\author{Nhat A. Nghiem }
\email{nhatanh.nghiemvu@stonybrook.edu}
\affiliation{Department of Physics and Astronomy, State University of New York at Stony Brook, Stony Brook, NY 11794-3800, USA}
\author{Tzu-Chieh Wei}
\affiliation{Department of Physics and Astronomy, State University of New York at Stony Brook, Stony Brook, NY 11794-3800, USA}
\affiliation{C. N. Yang Institute for Theoretical Physics, State University of New York at Stony Brook, Stony Brook, NY 11794-3840, USA}

\begin{abstract}
Following the celebrated quantum algorithm for solving linear equations (so-called HHL algorithm), Childs, Kothari and Somma [SIAM Journal on Computing, {\bf 46}: 1920, (2017)] provided an approach to solve a linear system of equations with exponentially improved dependence on precision. In this note, we aim to complement such a result for applying a matrix to some quantum state,  based upon their  Chebyshev polynomial approach. A few examples that motivate this application are included and we further discuss an application of this improved matrix application algorithm explicitly with an efficient quantum procedure.
\end{abstract}
\maketitle
\section{Main Procedure}
Quantum algorithm for solving linear systems, so-called HHL, was originally introduced in Ref.\cite{harrow2009quantum}, in which the authors aimed to find solution to the following equations: $A \cdot x = B$, where $A$ is $N \times N$ matrix that is s-sparse ($s \ll N$), and $B$ is an $N \times 1$ unit vector. In the ideal case, $x$ could be written down explicitly as $x = A^{-1} \cdot B$. Hence, the ability to invert the matrix $A$ yields the solution directly. Given access to entries of $A$, and a procedure to prepare  the vector $B$ (or more precisely, $\ket{B}$), the outcome of HHL algorithm is a unit vector (i.e., quantum state) proportional to the actual solution: $\ket{x}  = {x}/{||x||}$ where $||x||$ is the length ($l_2$-norm) of the solution. Let $\kappa$ be conditional number of $A$ (which is the ratio between 
 the largest and smallest singular values). The running time of the  HHL algorithm is $poly (log(N), \kappa, s, \frac{1}{\epsilon}   )$, where $\epsilon$ is the error tolerance. \\

The error dependence $\mathcal{O}(1/\epsilon)$ is a result of the quantum phase estimation (QPE) subroutine used in HHL algorithm. This dependence has been significantly improved in~\cite{childs2017quantum} to scale as $\mathcal{O}(poly(log(1/\epsilon)))$. As mentioned in~\cite{childs2017quantum},  due to the finite-bit precision required to achieve an error $\epsilon$ eventually, the QPE subroutine in~\cite{harrow2009quantum} limits a better scaling of error dependence. Instead, the authors of~\cite{childs2017quantum} outlined two approaches based on Fourier and Chebyshev series decomposition.
These two methods are implemented differently, but they share the same striking property: their truncations at relatively low orders ($\sim \mathcal{O}(poly(log(1/\epsilon))$) guarantee an approximation error $\epsilon$ of the given function, i.e., ${1}/{x}$, for the matrix inversion. These series decompositions are then integrated into quantum algorithms thanks to a procedure to apply a linear combination of unitary operators. Details can be found in~\cite{childs2017quantum}, and, in particular, see  their Lemma 6 and Lemma 7. \\

As we have mentioned previously, the key to solve the linear equation is to invert a given matrix $A$, i.e., to implement the action of $A^{-1}$. While the application of the quantum matrix inversion method still continues to be explored,  for example, Ref.~\cite{clader2013preconditioned} provides a concrete case that benefits its quantum advantage,  many problems instead require the ability to \textit{apply} the matrix multiplication itself, so as to implement the multiplication of $A$ itself. In~\cite{wiebe2012quantum}, the authors consider the problem of data fitting in the quantum setting. Aside from inverting a matrix, their algorithm requires the ability to prepare a state that is proportional to $F \ket{B}$, where $F$ is some sparse matrix of a  given size and $\ket{B}$ is some given unit vector. Another example is the quantum power method executed by consecutive application of a given matrix to some input state~\cite{nghiem2022quantum}, combined with Hadamard test to obtain the largest eigenvalue of a given matrix.\\

As worked out in \cite{wiebe2012quantum}, multiplication of a matrix is an adaptive version of HHL algorithm \cite{harrow2009quantum}. After the QPE step, one simply performs controlled-rotation on the ancilla with the multiplied eigenvalues (instead of dividing them) to achieve the goal of multiplication (see~\cite{harrow2009quantum, wiebe2012quantum} for details). Since it is just a slight modification of the HHL algorithm, the QPE subroutine is still employed and hence, the running time still scales as $\mathcal{O}(1/\epsilon)$. It is very natural to ask if such an error scaling could be improved to $\mathcal{O}(poly(log(1/\epsilon)))$, as for 
 the matrix inversion case. The answer is affirmative, and it is surprisingly simple to achieve this, as we explain below.  \\

We now discuss the Chebyshev series approach in~\cite{childs2017quantum}. The two main ingredients of this approach are:  (1) approximating a function (in the matrix inversion case, it is 1/$x$) by a linear combination of Chebyshev polynomials, and (2) implementing each Chebyshev polynomial by a quantum walk~\cite{berry2015hamiltonian}. In~\cite{childs2017quantum}, all these ingredients are discussed comprehensively, and hence we would not repeat the details here. Instead, we recapitulate a key result from there. For completeness, a brief review of key ingredients from~\cite{childs2017quantum} is provided in Appendix~\ref{sec: review}. 

\begin{lemma}
    \cite{childs2017quantum} Let A be a symmetric, $N \times N$, s-sparse matrix with eigenvalues in the range $[\frac{1}{\kappa}, 1]$. Define $U_n (A)$ as following: 
    \begin{align*}
        U_n (A) \ket{0^m}\ket{\phi} = \ket{0^m}T_n(A/s) \ket{\phi} + \ket{\Phi_{\perp}}
    \end{align*}
    where $m = log(2N) +1$, $ \ket{\Phi_{\perp}}$ is a (unnormalized) state that is orthogonal to $\ket{0^m}T_n(A/s) \ket{\phi}$: $ \ket{0^m}\bra{0^m} \otimes \mathbb{I} \ket{\Phi_{\perp}} = 0$,  and $T_n$ is the $n$-th Chebyshev polynomial of the first kind. Given coherent access to entries of A, then $U_n$ could be implemented up to error $\epsilon$ in time 
    \begin{align*}
        \mathcal{O}\big( n( log N + log^{2.5} (\kappa s n/ \epsilon) )   \big).
    \end{align*}
    \label{lemma: randomwalk}
    using $\mathcal{O}(n)$ queries to entries of A.
\end{lemma}

Recall that in the above Lemma, by coherent access to entries of matrix $A$, we mean that there exists a procedure $P_A$ that returns the entries of $A$ given entries index, i.e:
\begin{align}
    P_A \ket{i}\ket{j}\ket{ \bf{0} } \rightarrow \ket{i}\ket{j}\ket{a_{ij}}
\end{align}

where the third register holds the encoded values of corresponding entries $a_{ij}$. We emphasize that this is a very standard assumption in the context of Hamiltonian simulation \cite{berry2007efficient,berry2012black,berry2015hamiltonian}. In \cite{harrow2009quantum}, the assumption is slightlly modified as they assume entries of given matrix is row-computable, i.e, they can be computed on demand. \\

We remind that Chebyshev polynomials of the first kind can be obtained from the recurrence relation as follows. 
\begin{align*}
    T_0 (x) &= 1, \\
    T_1 (x) &= x, \\
    T_{n+1}(x) &= 2x T_n(x)  - T_{n-1}(x).
\end{align*}
Whereas in the matrix inversion case, we need to approximate 1/$x$ by a Chebyshev series with complicated coefficients~\cite{childs2017quantum}, in the matrix multiplication case, it is simply that $T_1(x) = x$. This means that we only need to implement $U_1(A)$ in order to successfully multiply  a given state $\ket{\phi}$ by a matrix. In other words, the representation of function $x$ as a Chebyshev series is exact and extremely short. It further means that no approximation error is induced. The only source of error comes from the implementation of $T_1(A)$, which is done by using quantum walk. Now we state our main result. 

\begin{theorem}[Matrix Multiplication]
\label{thm: matrixapply}
Let $\ket{B}$ be some quantum state of dimension $N \times 1$, and $U_B$ denotes the unitary circuit that prepares $\ket{B}$. Let A be a symmetric matrix of size $N\times N$, s-sparse and have  a conditional number $\kappa$. Given cohrent access to matrix A, then there exists a quantum procedure that prepares the following state
\begin{align*}
    \ket{\phi} = \frac{ A\ket{B} }{||A\ket{B}||}
\end{align*}
in time 
$$\mathcal{O}\big( poly(log N, log(\kappa), s, log(\frac{1}{\epsilon}) ) \big). $$
\end{theorem}

The proof is straightforward. We first apply $I^m \otimes U_B $ to $\ket{0}^m  \otimes \ket{0}^{log N}$ to obtain $\ket{0}^m \ket{B}$ (the tensor product sign is omitted here) and  then apply the result of Lemma~\ref{lemma: randomwalk} with $n$ = 1 to obtain the following state
\begin{align}
    U(A) \ket{0}^m \ket{B} = \ket{0}^m \frac{A}{s} \ket{B} + \ket{\Phi_{\perp}},
\end{align}
where we have used the fact that $T_1(A) = A$. Now we measure the first $m$ qubits, post-select on obtaining $\ket{0}^m $ to obtain the desired state $\ket{\phi}$. The measurement succeeds with a probability 
$ ||A\ket{B}||^2/s^2 $
which can be improved to near unity by quantum amplitude amplification~\cite{brassard2002quantum}, at the cost of inflicting the circuit depth by a factor $\mathcal{O}( ||A\ket{B}||/s )$. The total running time is thus $\mathcal{O}( T_U s/ ||A\ket{B}||)$,  where $T_U$ is the time required to implement $U_1(A)$. By plugging in $T_U$ the expression from Lemma \ref{lemma: randomwalk}, we obtain the final running time as stated. Note that we have absorbed the length of solution $||A\ket{B}||$ into the running time. \\

Before we head into some applications, we make a quick comparison between our improved matrix multiplication algorithm versus previously known method for multiplying matrix, which occurs in the context of data fitting \cite{wiebe2012quantum}. For completeness, the statement of the matrix multiplication problem is, for a given state $\ket{b}$, we need to prepare the state $A\ket{b}/ |A\ket{b}|$ where $|.|$ refers to the norm.  In their work \cite{wiebe2012quantum}, the authors essentially adapt HHL algorithm \cite{harrow2009quantum} to implement matrix multiplication by simply adjusting the phase rotation step, which produces the desired state in time 
\begin{align*}
    \mathcal{O}]\Big( \log(N) s \kappa^2 / \epsilon  \Big) 
\end{align*}

Meanwhile, as we pointed out in Thm \ref{thm: matrixapply}, the desired state could be produced in time that scales polylogarithmically to most of parameters, except sparsity $s$. It demonstrates the quite surprising power of the Chebyshev approach that was first employed in \cite{childs2017quantum}. Furthermore, in this approach, only a single query to entries of $A$ is being used.

\section{Applications}
In this section, we highlight some interesting applications of the above simple procedure for matrix multiplication. We will soon appreciate a very remarkable efficiency that the above method yields.

\subsection{Accelerating Eigenvalues Finding}
As we have mentioned from the beginning, one of the problem that motivates this work is quantum power method~\cite{nghiem2022quantum}. More specifically, a similar technique as in \cite{harrow2009quantum, wiebe2012quantum} was used to consecutively `apply' a given matrix $A$ to some initial state $\ket{x_0}$. By `apply', we mean the unitary $U_A$ that acts as following:
\begin{align}
    U_A\ket{0}\ket{x_0}  = \ket{0}A\ket{x_0} + \ket{1}\ket{Garbage},
\end{align}
where $\ket{Garbage}$ is some unimportant state (not properly normalized). Combining such a procedure with the Hadamard test, the largest eigenvalue of $A$ can be estimated. As mentioned in Theorem 1 of~\cite{nghiem2022quantum}, the running time of the quantum power method is 
\begin{align}
    \mathcal{O}\Big( \frac{ \sqrt{n} \log(n) s \kappa^2}{\epsilon^4} \Big),
    \label{eqn: quantumrunningtime}
\end{align}
where $n$ is the dimension of $A$, $s$ is the sparsity of $A$, $\kappa$ is the conditional number of $A$, $k$ is the number of iterations, and $\epsilon$ is the error tolerance (the additive approximation to the true largest eigenvalue).\\

While the speedup with respect to $n$ is remarkable, we shall see that the improved matrix application~\ref{thm: matrixapply} can improve the scaling on $\kappa$ and $\epsilon$. The crucial step in such a quantum power method is the HHL-like procedure, where the eigenvalues of $A$ are extracted via quantum phase estimation, followed by rotation of ancilla controlled by the phase register. As can be seen from Theorem~\ref{thm: matrixapply}, matrix multiplication can be done very efficiently with the Chebyshev method \cite{childs2017quantum}. The dependence on dimension $n$ and conditional number $\kappa$ and even tolerance error $\epsilon$ are all polylogarithmic. With such an improvement in performing matrix multiplication, we establish the following faster routine for quantum power method:
\begin{theorem}
Given coherent access to a Hermitian matrix A, its largest eigenvalue could be estimate to additive accuracy $\epsilon$ with running time
\begin{align}
   \mathcal{O} \Big(  (log(n) + log^{2.5}( \frac{\kappa s}{\epsilon} )) \frac{s \kappa \sqrt{n}}{\epsilon^3}   \Big).  
\end{align}
\end{theorem}

For simplicity purpose, if we neglect the contribution of logarithmic terms (compared to the polynomial terms), then we the faster QPW method has the same scaling on dimension $n$ and sparsity $s$, almost quadratically better dependence on $\kappa$ and a minor improvement on error $\epsilon$.

\subsection{Estimating Trace of a Matrix}
We remark that the adaptive method in~\cite{wiebe2012quantum} has the same running time as the original HHL~\cite{harrow2009quantum}, which is quadratic in $\kappa$ and polynomial in sparsity $s$. In our aforementioned result, the dependence on $\kappa$ and $s$ are polylogarithmic, which is thus  highly efficient. It demonstrates the surprising utility of the Chebyshev series method introduced in~\cite{childs2017quantum}. As we have mentioned three examples above that rely on matrix multiplication, we foresee that matrix multiplication will be employed in other algorithms. The polylogarithmic dependence on all factors highlights that even if it is being used as 
 a subroutine in a larger algorithm, the cost induced would be very modest. To exploit further the surprising efficiency of the above method, we now provide an alternative solution to the trace estimation problem of some given matrix. This problem has occured in~\cite{rebentrost2014quantum}, as estimating trace of kernel matrix is a necessary step of building (quantum) support vector machine. Additionally, this problem is also  finds its  application the DQC-1 model of quantum computation~\cite{knill1998power,shor2008estimating}, which can, e.g., efficiently estimate the values of Jones polynomials. We shall see that, in a similar fashion to the Chebyshev method \cite{childs2017quantum}, our method to estimate trace of a matrix would not make use of Hamiltonian subroutine, e.g., as in~\cite{rebentrost2014quantum}. 

The full statement of trace estimation problem is straightforward as it seems. Suppose we are given access, for example, to an efficient procedure that computes entries of some Hermitian matrix $A$ with the conditional number $\kappa$ and sparsity $s$, and the goal is to estimate $\Tr(A)$. Without loss of generalization, assume that dim($A$) = $N$ is a power of 2, i.e, $N=2^n$ where $n$ is some positive integer. Classically this problem is very simple, as we only need to perform the summation of diagonal entries. Now we sketch our new quantum solution that employs matrix multiplication as a subroutine. 

Similar to~\cite{childs2017quantum, harrow2009quantum}, we assume that entries $a_{ij}$ of $A$ is accessible via a procedure denoted as $P_A$. What we do next is to `relocate' those diagonal entries of $A$ to the first column by performing the following trick. Remind that the operation of $P_A$ is done as follows: 
$$ P_A\ket{i}\ket{j}\ket{00..00} = \ket{i}\ket{j}\ket{a_{ij}}.$$
where the number of $0$ on the third register on the left hand side of the above equation is the necessary number of bits required to store the corresponding entries. We note that the diagonal entries are $\{ a_{ii} \}$ and entries on the first column are $\{ a_{i0}\}$. Before we use $P_A$ to query the index, we add CNOT layers, denoted as C. Such CNOT layers C acts on the diagonal index as: 
$$ \ket{i}\ket{i} \rightarrow \ket{i}\ket{0}, $$
and 
$$ \ket{i}\ket{0} \rightarrow \ket{i}\ket{i}.$$
Then reverse the CNOT layers to return the old index. We obtain the following: 
\begin{align}
   P_A' \ket{i}\ket{i} = C^\dagger P_A C \ket{i}\ket{i} = \ket{i}\ket{i}\ket{a_{i0}},
\end{align}
and 
\begin{align}
 P_A' \ket{i}\ket{0} = \ket{i}\ket{i} \ket{a_{ii}}.
\end{align}
Effectively, the modified blackbox $P_A'$ would  `return' the new matrix that has entries in a different order compared to $A$. However, the first column of, say, $A'$ is exactly the diagonal entries of $A$. We remark that what we describe above is just a simple way to relocate the entries of given matrix $A$. Basically, one can opt to relocate, or permute entries of $A$ in different ways, as long as the first column of the new matrix is the diagonal of $A$.  \\

Now we are ready to describe our main algorithm. We note the following simple matrix multiplication identity:
\begin{align*}
\begin{bmatrix}
 m_{00} & m_{01} & \cdots & m_{0,n-1} \\
 m_{10} & m_{11} & \cdots & m_{1,n-1} \\
 \vdots & \\
 m_{n-1,0} & m_{n-1,1} & \cdots & m_{n-1,n-1}.
\end{bmatrix}
\times 
\begin{bmatrix}
1  \\
0 \\
\vdots \\
0
\end{bmatrix}
= 
\begin{bmatrix}
m_{00} \\
m_{10} \\
\vdots \\
m_{n-1,0} 
\end{bmatrix}.
\end{align*}

It means that if we simply apply the matrix $A'$ to the first basis state $\ket{0^n}$ and it results in a vector $u$ with entries from the first column of $A'$, which are exactly diagonal entries of $A$. 
%
Let $v$ be an $N\times 1$  vector with all entries 1 (up to an overall normalization) then it is obvious that
\begin{align}
    \braket{v,u} = \sum_{i=0}^{N-1} a_{ii} = \Tr(A).
\end{align}
The corresponding normalized quantum state $\ket{v} = \frac{1}{\sqrt{N}} v$ can simply be prepared using Hadamard gates, i.e., $H^n \ket{0^n}$, with $N=2^n$. Therefore, in order to estimate $\Tr(A)$, we can employ the SWAP test algorithm~\cite{nielsen2002quantum} with the following two input states: \\

$\bullet$ The first state is (see Lemma \ref{lemma: randomwalk} for definition of $U(A')$ and $m = log(2N)+1 )$
\begin{align}
    \ket{\Phi_1} = U(A')\ket{0}^m \ket{0^n} = \ket{0}^m \frac{A'}{s} \ket{0^n} + \ket{\text{Garbage}},
\end{align}
where we note that $\ket{\text{Garbage}}  $ satisfies $\ket{0^m}\bra{0^m} \otimes I \ket{\text{Garbage}} = 0$; and $s$ is the sparsity of matrix A. \\

$\bullet$ The second state is
\begin{align}
    \ket{\Phi_2}  = \ket{0}^m \ket{v}.
\end{align}

Their inner product $\braket{\Phi_1,\Phi_2}$ is $\frac{1}{\sqrt{N}} \sum_{i=0}^{N-1} a_{ii} = \Tr(A)/ (s \sqrt{N})$. We remark that $\Tr(A)$ could be complex. Therefore, the real and complex part of $\Tr(A)$ can simply be estimated to an accuracy $\delta$, respectively, by running the Hadamard test circuit $\mathcal{O}(1/\delta^2)$ times. The whole process, including applying matrix $A'$ and estimating $\Tr(A)$, is highly efficient with respect to key parameters, which are dimension $N$, conditional number $\kappa$ and sparsity $s$. As we have seen that the matrix multiplication has running time that is polylogarithmically in all factors, i.e., $N$, $\kappa$, and $s$, therefore, the above trace estimation procedure can be applied to dense matrices with high conditional numbers. The running time of above quantum algorithm is 
\begin{align}
        \mathcal{O}\big( ( log N + log^{2.5} (\kappa s / \epsilon) ) / \delta^2  \big).
\end{align}
We have noted that this problem can be done easily using classical algorithm. We just simply need to query the matrix to get the diangonal entries and perform the summation, which can be done in logarithmic time using parallel algorithm. Therefore, classical computer can perform better in regards of time complexity. However, classical algorithm requires as much as N queries to matrix A in order to obtain diagonal entries. Meanwhile, in order to perform matrix multiplication, $\mathcal{O}(1)$ queries to matrix A is sufficient \cite{childs2017quantum}. We remark that the above quantum running time is highly efficient w.r.t all parameters $N,\kappa,s$, as long as when $\kappa$ and $s$ grows polynomially as a function of $N$. For completeness, we restate the above quantum algorithm in the following theorem with the condition that the matrix $A$ is not too ill-conditioned. 

\begin{theorem}
    Given coherent access to some Hermitian matrix A, then its trace can be estimated to additive accuracy $\delta$ in time 
    
    \begin{align*}
        \mathcal{O}\big( poly\log(N)/\epsilon^2 \big).
    \end{align*}
    
    where the highest order of $polylog(N)$ is about 2.5 
\end{theorem}

Another solution to the trace estimation problem which does not require relocation of entries is clearly possible. We first apply $H^{\log N} \otimes I^{\log N}$ to $\ket{0}^{\log N} \ket{0}^{\log N}$, following by a CNOT layer composed of $\log N$ CNOT gates, to obtain the following state:
\begin{align}
\ket{\phi} = \frac{1}{\sqrt{N}}\sum_{i=1}^N \ket{i}\ket{i}.
\end{align}

We now use Lemma \ref{lemma: randomwalk} to perform the following unitary $U_1 (A) \otimes I^{\log N}$ to the above state. We end up having:
\begin{align}
    \ket{\phi_1} = \frac{1}{\sqrt{N}} \sum_{i=1}^N \Big( \ket{0^m} (A/s) \ket{i} + \ket{Garbage}_i  \Big) \ket{i},
\end{align}
where $\ket{Garbage}_i$ is the redundant state that is orthogonal to $\ket{0}^m A\ket{i}$ for all $i$. We remark that $A\ket{i}$ is the $i$-th column of $A$. The diagonal entries of $A$ are now `encoded' in the state $A\ket{i}\ket{i}$. We now prepare the following state (we have mentioned a procedure previously):
\begin{align}
    \ket{\phi_2} = \frac{1}{\sqrt{N}} \ket{0^m} \sum_{i=1}^N \ket{i}\ket{i}.
\end{align}
The inner product of the above two states is
\begin{align}
    \braket{\phi_2|\phi_1} = \frac{1}{N} \sum_{i=1}^N \bra{i}\bra{i} A \ket{i}\ket{i} = \frac{1}{s N} \Tr(A).
\end{align}

By using Hadamard test, we can estimate the real and complex part of $\Tr(A)$, or more precisely, of $\Tr(A)/(sN)$. 

\subsection{Estimating Trace of Product of Two Matrices}
In the above method, we use the simple trick to relocate those diagonal entries of given matrix $A$. However, that trick is not  convenient to use when dealing with the problem of estimating $\Tr(A\cdot B)$ where $A$ and $B$ are two matrices of size $N\times N$ (we assume the same condition on the bound of their spectrum, as well as Hermitian property as in the previous section). Before diving into the quantum algorithm, we quickly review the computation of trace. Let $C = A\cdot B$, then 
\begin{align}
    \Tr(C) = \sum_{i}  \sum_j a_{ij} b_{ji},
\end{align}
where $a_{ij}$ (respectively $b_{ij})$ refers to entries of $A$ (and respectively $B$). Since A and B are generally complex, $\Tr(AB)$ could be complex. We note that the term 
\begin{align}
    \sum_j a_{ij} b_{ji} 
\end{align}
is exactly the inner product between $i$-th row of $A$ and $i$-th column of $B$. However, as we have assumed $A$ to be Hermitian, we know $i$-th row of $A$ is the same is $i$-th column of $A$ up to a complex conjugation. Our quantum algorithm begins with the following state $\ket{0^n}$. Applying a layer of Hadamard gates $H^n$, we obtain
\begin{align}
    \ket{\phi_0} = \frac{1}{\sqrt{N}} \sum_{i=0}^{N-1} \ket{i},
\end{align}
where $\{\ket{i}\}_{i=0}^{N-1}$ are computational basis states, and  we then append another register initialized in $\ket{0}^m \ket{0}^n$ (where $m= log(2N)+1 $), followed by applying layers of CNOT using $\{\ket{i} \}$ as the control bits and $\ket{0}^n$ as the target qubits. We arrive at the state
\begin{align}
    \ket{\phi_1} = \frac{1}{\sqrt{N}} \sum_{i=0}^{N-1} \ket{i}\ket{0}^m\ket{i}.
\end{align}
The reason why we need $\ket{0}^m$ is to perform matrix application subsequently (see Lemma \ref{lemma: randomwalk}). Now we perform the matrix application of $B$ to the second register, with the aim of obtaining $B\ket{i}$ for all $\ket{i}$, and we obtain
\begin{align}
   \ket{\phi_2} =  \frac{1}{\sqrt{N}} \sum_{i=0}^{N-1} \ket{i} \Big( \ket{0}^m \frac{B}{s_B} \ket{i} + \ket{G_i} \Big),
\end{align}
where $\ket{G_i}$ is abbreviation to garbage state as a result of matrix application to $\ket{i}$. Again, we emphasize that this state is unimportant. 

Now we make a simple observation: since $\ket{i}$ is the $i$-th basis state, the entries of $B\ket{i}$ is exactly the $i$-th column of $B$, denoted as $B^i$. Now we follow the same three steps as above, but instead of applying matrix $B$ at the end, we apply matrix $A^\dagger$ (it is easy to see that local access to $A$ yields local access to $A^\dagger$ as it is just a matter of transpose plus complex conjugation). We thus obtain the following state
\begin{align}
    \ket{\Phi_2} = \frac{1}{\sqrt{N}} \sum_{i=0}^{N-1} \ket{i} \Big( \ket{0}^m \frac{A^\dagger}{s_A} \ket{i} + \ket{G'_i} \Big).
\end{align}

where $s_A,s_B$ is the sparsity of A and B respectively. It is straightforward to see that $A^\dagger \ket{i}$ is the $i$-th column of $A^\dagger$, which is the $i$-th row of $A$ due to its Hermitian property. A further important observation is that, the inner product of the above two states is proportional to
\begin{align}
   \Big(\sum_{j=0}^{N-1} \bra{j}\bra{j} A \Big) \cdot \Big( \sum_{i=0}^{N-1} \ket{i} B\ket{i} \Big) = \Tr(AB).
\end{align}
So now the goal is to evaluate the above summation, using $\ket{\phi_2}$ and $\ket{\Phi_2}$. At first, a simple solution is to measuring those two states separately and post-selecting on $\ket{0}^m$, and we will be left with a (normalized) state corresponding to $\sum \ket{i}A\ket{i}$. The renormalization factor can be revealed via measurement process. We then plug those post-selected state to the Hadamard test circuit. However, this method will induces  further cost. We now aim to incorporate such measuring/post-selecting step so as to save some computational time. Typically, in the Hadamard test, one measures the first qubit of the circuit and based on measurement probability to estimate the real/complex part of inner product (more precisely of given two states. Here, we put extra measurements on those $m$ registers of both states $\ket{\phi_2}$ and $\ket{\Phi_2}$. By post-selecting the measurement outcome to be $\ket{0}$ on the first register, and $\ket{0}^m$ on both registers of $\ket{\phi_2}$ and $\ket{\Phi_2}$ and estimate the probability outcome, we then can estimate the following quantities that correspond to real and imaginary part of $\Tr(AB)$, respectively:
\begin{align}
    \frac{\Re (\Tr(AB))}{N s_A s_B }  \text{ and } \frac{\mathcal{I} (\Tr(AB))}{N s_A s_B}
\end{align}
to accuracy $\delta$ using $\mathcal{O}(1/\delta^2)$ repetitions. Hence, $|\Tr(AB)|$ could also be estimated. We state the following theorem.
\begin{theorem}
Given access to entries of two matrix $A$ and $B$, the trace of $AB$ could be estiamted to additive accuracy $\delta$ in time
    \begin{align}
    \mathcal{O}\Big( poly(\log N,  \log \kappa), 1/\delta^2 \Big)
\end{align}
\end{theorem}

In order to compute $\Tr(AB)$, matrix multiplication needs to be done before summing over the diagonal entries. There are many classical algorithms for matrix multiplication, but none are known for achieving such task within polylogarithmic time. Therefore, the quantum algorithm outlined above yields exponential speedup with respect to dimension N of given matrix. 

\subsection{Estimating Frobenius Norm}
Given a Hermitian matrix $A$ of size $N \times N$, its Frobenius norm is computed as:
\begin{align}
    |A|_F = \sqrt{ \sum_{i=1}^N \lambda_i^2 },
\end{align}
where $\{\lambda_i\}$ are eigenvalues of $A$ (note that since $A$ is Hermitian then all eigenvalues are real). Revealing the full spectrum of $A$ is apparently a daunting task. However, by definition, we have:
\begin{align}
    |A|_F = \sqrt{\Tr(A^\dagger A)}.
\end{align}
Hence, as straightforward as it seems, the square of Frobenius norm is a direct result of previous section, since we are able to apply the matrix $A$ and $A^\dagger$. The running time is similar to equation (23). \\

As an alternative solution, now we describe how to compute such Frobenius norm with matrix application technique combined with quantum phase estimation. We first prepare the completely mixed state:
\begin{align}
    \rho_I = \frac{1}{N} \sum_{i=0}^{N-1} \ket{i}\bra{i},
    \label{eqn: mix}
\end{align}
which can be prepared simply by applying $H^n$ to $\ket{0}^n $ ($n = log(N)$) to obtain
\begin{align}
    \frac{1}{\sqrt{N}} \sum_{i=0}^{N-1} \ket{i},
\end{align}
Then apply the CNOT layers to obtain: 
\begin{align}
    \frac{1}{\sqrt{N}} \sum_{i=0}^{N-1} \ket{i}\ket{i}.
\end{align}

Tracing out the second register we would obtain the desired mixed state $\rho_I$. The algorithm now proceeds simply with the application of Lemma \ref{lemma: randomwalk} with mixed state input $\ket{0}^m \bra{0}^m \otimes \rho_I$ instead of pure state. The output of such application is then:
\begin{align}
    \ket{0}^m \bra{0}^m (A/s_A) \rho_I (A^\dagger/s_A) + \rho_{garbage},
\end{align}
where $\rho_{garbage}$ is the part that is orthogonal to $\ket{0}^m \bra{0}^m$ under trace. 
Now we perform measurement on the first m register and post-select on $\ket{0}^m$. The probability of such measurement is 
$$ \Tr ( A\rho_I A^\dagger  ) / s_A^2 = \frac{ \sum_{i=1}^N \lambda_i^2 }{N s_A^2 }$$
Due to the fact that the maximally mixed state Eqn.~(\ref{eqn: mix}) is diagonal in the space spanned by eigenvectors of $A$ (with coefficient $1/N$). In such basis, $A$ is diagonal (with entries $\{\lambda_i\}$ on the diagonal). Hence, simple algebraic multiplication yields the above probability. In order to estimate such probability to accuracy $\delta$, $\mathcal{O}(1/\delta^2)$ repetitions are sufficient. Since the preparation of mixed state $\rho_I$ is highly efficient, taking almost a single step, hence we have the following result: 
\begin{theorem}
Given access to matrix $A$, its Frrobenius norm $|A|_F$ could be estimated to additive accuracy $\delta$ in time: 
    \begin{align}
    \mathcal{O}\Big( poly(\log N,  \log \kappa), 1/\delta^2 \Big).
\end{align}
\end{theorem}

\section{Discussion and Conclusion}

Motivated by some recent progress in quantum algorithms, such as~\cite{childs2017quantum, harrow2009quantum, nghiem2022quantum, wiebe2012quantum}, we point out a simple way to apply the techniques developed in~\cite{childs2017quantum} to a broader context. The running time of such matrix application techniques is surprisingly efficient. Aside from significantly  accelerating the data fitting algorithm of Ref.~\cite{wiebe2012quantum} with our approach, in particularly, we provide a new way to deal with trace estimation problem, which might be applied to the DQC-1 algorithm~\cite{knill1998power,shor2007estimating}. Given that a huge amount of problems in engineering and physics require the ability to manipulate matrices, including performing their multiplication, we suspect that the simple method outlined in this work can find many applications. To leverage the result regarding estimating the trace of a matrix or product of two matrices as we describe above, we would like to mention the so-called Frechet distance problem, which aims to measure the difference between probability distributions. Suppose two multivariate Gaussian distributions with means $\mu_X$ and $\mu_Y$ and covariance matrices $C_X$ and $C_Y$ are given, the Frechet distance between these distributions is defined as
$$ d^2 = |\mu_X-\mu_Y|^2 + \Tr\Big( C_X + C_Y - 2 \sqrt{C_X C_Y}  \Big). $$
While the estimation of $\Tr(C_X)$ and $\Tr(C_Y)$ are straightforward, the last term $\Tr(\sqrt{C_X C_Y})$ is a bit more tricky as our above method requires further local access to $\sqrt{C_X}$ and $\sqrt{C_y}$. If, instead of $A$ and $B$, we have oracular access to their square root, then the solution is clear as we just need to apply our above algorithm. We would not discuss in greater detail about such oracle construction here as it is not the main scope of our work. A more interesting question is raised here, that whether or not we can estimate $\Tr(\sqrt{AB})$ given local access to $A$ and $B$ only? In principle, it could be done by using a general singular value decomposition (in this case if $A$ and $B$ are Hermitian, then it would be eigenvalues), adopting HHL technique~\cite{harrow2009quantum}. However, HHL algorithm employs quantum phase estimation as a subroutine and hence, would likely to incur $\mathcal{O}(1/\epsilon)$ time ahead, meanwhile our goal for this particular problem, which is also the goal of the paper, is to reduce such running time to polylogarithmically on the error-tolerance. We leave such a problem for future work.

\section*{Acknowledgements}
This work was supported in part  by the U. S. Department of Energy, Office of Science, National Quantum Information Science Research Centers, Co-design Center for Quantum Advantage (C2QA)
under contract number DE-SC0012704. We also acknowledge the support from a Seed Grant from Stony Brook University's Office of the Vice President for Research.

\section*{Declarations}
On behalf of all authors, the corresponding author states that there is no conflict of interest. 

\section*{Data Availability Statement}
There is no data generated in this work.

\bibliography{ref.bib}{}
\bibliographystyle{unsrt}

\clearpage
\newpage
\onecolumngrid

\appendix
\section{Review of Chebyshev Approach}
\label{sec: review}
Here we make a review of Chebyshev approach that was employed in \cite{childs2017quantum}, which is essentially built upon quantum walk technique \cite{berry2012black, berry2015hamiltonian}. What we will describe below is more or less a summary of Section 4 in Ref.~\cite{childs2017quantum}, the result of which was used in our main text. \\

Let $A$ be a $d$-sparse ($s$ was used in the main text) $N \times N$ Hermitian matrix and a procedure, or black-box that could query the entries of $A$. Define the following state on the Hilbert space $\mathcal{C}^{2N} \otimes \mathcal{C}^{2N}$:
\begin{align}
    \ket{\psi_j} = \ket{j} \otimes \frac{1}{\sqrt{d}} \sum_{k \in [N]; A_{ij} \neq 0} \Big( \sqrt{A_{ij}^*} \ket{k} + \sqrt{1- |A_{ij}|} \ket{k+N}  \Big)
\end{align}
We also define the following isometry $T$ from $\mathcal{C}^N$ to $\mathcal{C}^{2N} \otimes \mathcal{C}^{2N}$:
\begin{align}
    T = \sum_{j\in [N]} \ket{\psi_j}\bra{j}.
\end{align}
We note that $T$ is not unitary, and, therefore, what we would need is a unitary version of $T$, which is $U_T$ acting as following:
\begin{align}
    U_T \ket{0^m}\ket{\phi} = T\ket{\phi},
\end{align}
for some $\ket{\phi} \in \mathcal{C}^N$ and $m = log(2N) +1$.

The so-called \textit{walk operator} is defined as: 
\begin{align}
    W = S(2T T^\dagger - I),
\end{align}
where $S$ is the SWAP operator on the space $\mathcal{C}^{2N} \otimes \mathcal{C}^{2N}$, i.e., $S \ket{j,k} = \ket{k,j}$. The implementation of $W$ (and $U_T$) was explicitly described in Lemma 10 of Ref.~\cite{berry2015hamiltonian}. We now summarize the structural property of $W$ that makes it highly beneficial for many quantum algorithms, including the improved quantum linear solver~\cite{childs2017quantum}, and quantum simulation~\cite{berry2015hamiltonian}.  \\

Let $\ket{\lambda}$ and $\lambda$ be eigenvector and eigenvalue of $A/d$ (note that the scaling by $d$ doesn't have further systematic problem, as the spectrum remains the same, only eigenvalues got scaled by a factor). 
Within the subspace spanned by $T\ket{\lambda}$ and $ST\ket{\lambda}$,  $W$ admits the following block form:
\begin{align}
    \begin{bmatrix}
        \lambda & -\sqrt{1-\lambda^2} \\
        \sqrt{1-\lambda^2} & \lambda
    \end{bmatrix}.
\end{align}
The proof can be found in Lemma 15 of \cite{childs2017quantum}. The above form of $W$ possess the following remarkable property (Lemma 16 of \cite{childs2017quantum}),
\begin{align}
    W^n = \begin{bmatrix}
        T_n(\lambda) & -\sqrt{1-\lambda^2} U_{n-1}(\lambda) \\
        \sqrt{1-\lambda^2} U_{n-1}(\lambda ) & T_n(\lambda)
    \end{bmatrix},
\end{align}
where $T_n$ is the $n$-th Chebyshev polynomial of the first kind, and $U_n$ is the $n$-th Chebyshev polynomial of the second kind. The construction of $W_n$ is simply product of $n$ times application of $W$. Therefore, we can construct the unitary $U = U_T^\dagger W^n U_T$ that acts on $\ket{0^m}\ket{\phi}$ as follows,
\begin{align}
    U\ket{0^m}\ket{\phi} = \ket{0^m} T_n (A) \ket{\phi} + \ket{\Phi_{\perp}},
\end{align}
where $\ket{\Phi_{\perp}}$ is orthogonal to $\ket{0^m} T_n (A) \ket{\phi}$. As we have remarked in the main text, the 1st Chebyshev polynomial of the first kind $T_1(x)=x$ is sufficient for matrix multiplication purpose, which requires a single execution of $W$.

\end{document}